\documentclass[aps,pre,twocolumn,showpacs]{revtex4}
\usepackage{graphics,graphicx,subfigure}
\usepackage{amsmath}

\newcommand{\ben}{\begin{enumerate}}
\newcommand{\een}{\end{enumerate}}
\newcommand{\bit}{\begin{itemize}}
\newcommand{\eit}{\end{itemize}}
\newcommand{\be}{\begin{equation}}
\newcommand{\ee}{\end{equation}}
\newcommand{\bdm}{\begin{displaymath}}
\newcommand{\edm}{\end{displaymath}}
\newcommand{\bea}{\begin{eqnarray}}
\newcommand{\eea}{\end{eqnarray}}
\newcommand{\f}[1]{\fbox}

\begin{document}

\title{Measurability of vacuum fluctuations and dark energy}

\author{Christian Beck}
\affiliation{ School of Mathematical Sciences \\ Queen Mary,
University of London \\ Mile End Road, London E1 4NS, UK}
\email{c.beck@qmul.ac.uk}
\homepage{http://www.maths.qmul.ac.uk/~beck}
\author{Michael C. Mackey}
\affiliation{Centre for Nonlinear Dynamics in Physiology and
Medicine \\ Departments of Physiology, Physics and Mathematics \\
McGill University, Montreal, Quebec, Canada}
\email{mackey@cnd.mcgill.ca}
\homepage{http://www.cnd.mcgill.ca/people_mackey.html}
\altaffiliation{ also: Mathematical Institute, University of Oxford,
24-29 St Giles', Oxford OX1 3LB, UK}

\date{\today}

\vspace{2cm}

\begin{abstract}
Vacuum fluctuations of the electromagnetic field induce current
fluctuations in resistively shunted  Josephson junctions that are
measurable in terms of a physically relevant power spectrum. In
this paper we investigate under which conditions vacuum
fluctuations can be gravitationally active, thus contributing to
the dark energy density of the universe. Our central hypothesis
is that vacuum fluctuations are {\em gravitationally active} if
and only if they are {\em measurable} in terms of a physical
power spectrum in a suitable macroscopic or mesoscopic detector.
This hypothesis is consistent with the observed dark energy
density in the universe and offers a resolution of the
cosmological constant problem. Using this hypothesis we show that
the observable vacuum energy density $\rho_{vac}$ in the
universe is related to the largest possible critical temperature
$T_c$ of superconductors through $\rho_{vac}=\sigma \cdot
\frac{(kT_c)^4}{\hbar^3 c^3}$, where $\sigma$ is a small constant
of the order $10^{-3}$.
This relation can be regarded as an analog of the
Stefan-Boltzmann law for dark energy.
Our hypothesis is testable in  Josephson
junctions where we predict there should be a cutoff in the
measured spectrum at $1.7$ THz if the hypothesis is true.

\end{abstract}


\date{\today}

\maketitle

\section{Introduction}

A fundamental problem relevant to quantum field theories,
cosmology, and statistical
mechanics is whether vacuum fluctuations  are `real' in the sense
that the corresponding vacuum energy has a gravitational effect
\cite{jaffe, zeldo, schwinger, saunders}. Quantum electrodynamic
(QED) explanations of  phenomena such as the Casimir effect, the
Lamb shift, van der Waals forces, spontaneous emission from
atoms, etc. provide indirect evidence for the existence of vacuum
fluctuations. However, Jaffe \cite{jaffe} emphasizes that the
Casimir effect, which is generally believed to provide stringent
evidence for the existence of QED vacuum fluctuations, can be
explained without any reference to vacuum fluctuations. Indeed it
seems that for most QED phenomena vacuum fluctuations are a
useful mathematical tool to theoretically describe these effects,
but there are formulations such as Schwinger's source theory
\cite{schwinger} that completely avoid any reference to vacuum
fluctuations.

It is well known that the amount of vacuum energy formally predicted
by quantum field theories is too large by many orders of magnitude
if gravitational coupling is taken into account. While the currently
observed dark energy density in the universe is of the order
$m_\nu^4$, where $m_\nu$ is a typical neutrino mass scale, quantum
field theories predict that there is  an infinite vacuum energy
density, since the corresponding integrals diverge. Assuming a
cutoff on the order of the Planck scale, one still has a vacuum
energy density too large by a factor $10^{120}$. This is the famous
cosmological constant problem
\cite{weinberg,peebles,sahni,padman1,copeland,beck04}.

The relevant question would thus seem to be not whether vacuum
fluctuations exist (they certainly exist as a useful theoretical
tool), but under which conditions they have a physical reality in
the sense that they produce a directly measurable spectrum of
fluctuations in macroscopic or mesoscopic detectors which could
have a gravitational effect \cite{beck-mackey, frampton,
kiefer,bm2}. With respect to this question, a very interesting
experiment was performed by Koch, van Harlingen and Clarke
\cite{koch} in 1982. Koch et al. experimentally measured the
spectral density of  the noise current in a resistively shunted
Josephson junction and showed that the data were described  by
the theoretically derived spectrum
\cite{welton,senitzky,koch80theory,gardiner,kogan,levinson}
\begin{eqnarray}
 S_{I}(\omega)&=&\frac{2\hbar\omega}{R}\coth
\left( \frac{\hbar\omega}{2kT} \right) \nonumber \\ &=&\frac{4}{R}
\left[ \frac{1}{2}\hbar\omega+\frac{\hbar\omega}{\exp
(\hbar\omega/kT)-1} \right], \label{eq:power}
\end{eqnarray}
where $R$ is the shunting resistance, $\omega = 2 \pi \nu$ is the
frequency, $h$ is Planck's constant, $k$ is Boltzmann's constant,
and $T$ is the temperature.  The first term in this equation is
independent of temperature and increases linearly  with the
frequency $\nu$. This term is induced by zero-point fluctuations of
the electromagnetic field, which  produces measurable noise currents
in the Josephson junction \cite{gardiner}. The second term is due to
ordinary Bose-Einstein statistics and describes thermal noise. The
experimental data of \cite{koch}, reproduced in Figure 1, confirm
the form (\ref{eq:power}) of the spectrum up to a frequency of
approximately $\nu \simeq 6 \times 10^{11}$ Hz.

\begin{figure}[ht]
\includegraphics[scale=0.3]{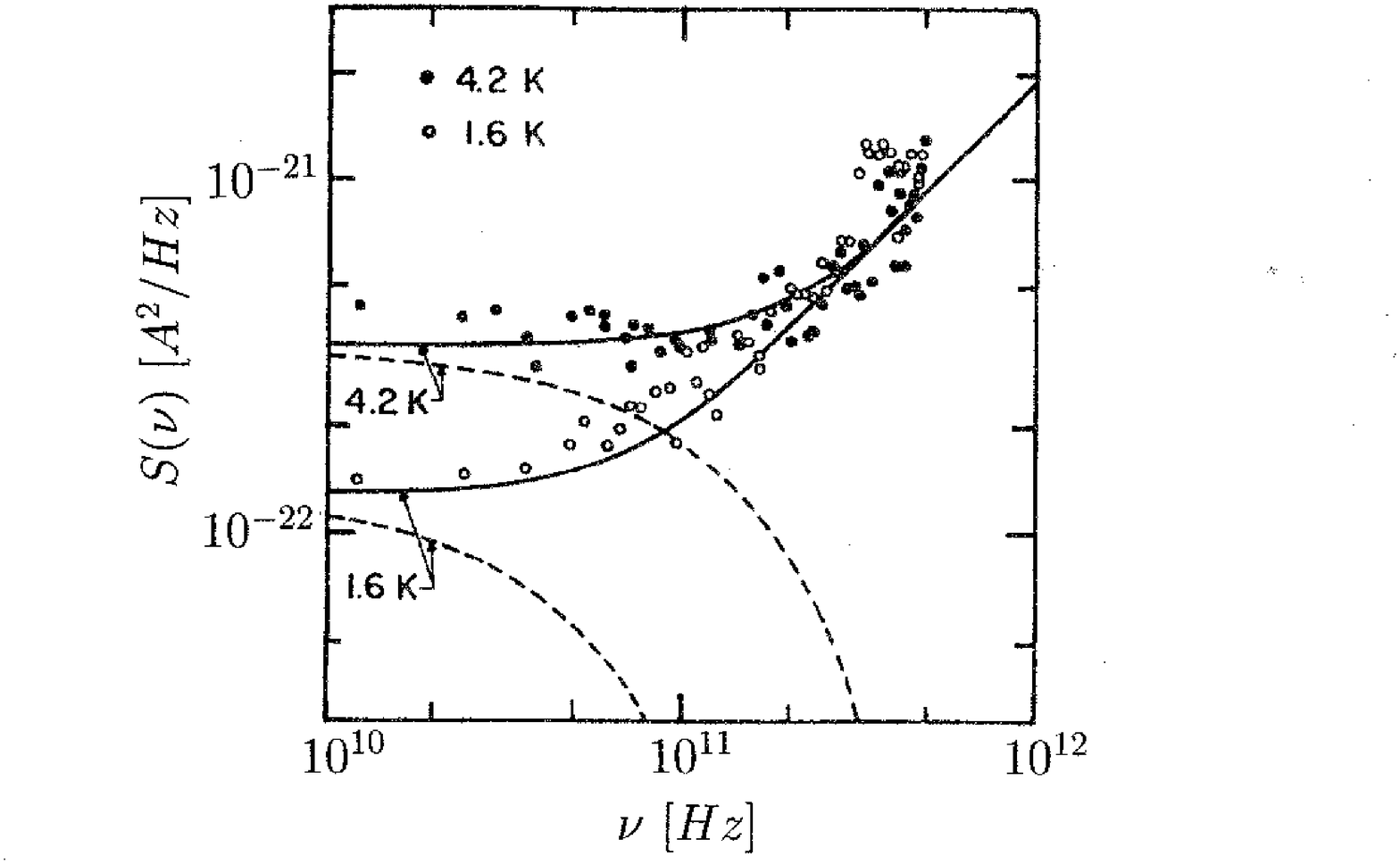}
\caption{Spectral density of current noise as measured in
\cite{koch} for two different temperatures. The solid line is the
prediction of eq.~(\ref{eq:power}), while the dashed line neglects
the zero-point term.} \label{fig:data}
\end{figure}

The  data in Fig.~1 represent a {\it physically measured} spectrum
induced by vacuum fluctuations.   The spectrum is measurable due to
subtle nonlinear mixing effects in Josephson junctions. These mixing
effects, which  have nothing to do with either the Casimir effect or
van der Waals forces, are a consequence of the AC Josephson effect
\cite{josephson, tinkham, landau2}.

In \cite{beck-mackey} we suggested an extension of  the Koch
experiment to higher frequencies. We based this suggestion on the
observation that if the vacuum energy associated with the measured
zero-point fluctuations in Fig.~1 is gravitationally active (in
the sense that the vacuum energy is the source of dark energy),
then there must be  a cutoff in the measured spectrum at a
critical frequency $\nu_c$. Otherwise the corresponding vacuum
energy density would exceed the currently measured dark energy
density \cite{bennett, spergel, spergelnew} of the universe. The
relevant cutoff frequency was predicted in \cite{beck-mackey} to
be given by
\begin{equation}
    \nu_c \simeq (1.69 \pm 0.05) \times 10^{12} \quad \mbox{Hz},
\label{eq:cutoff}
\end{equation}
only 3 times larger than the largest frequency reached in the 1982
Koch et al. experiment \cite{koch}.

In this paper we discuss  the measurability of vacuum fluctuations
as inspired by the Koch et al. experiment. We are  motivated by
the fact that this  experiment will now be repeated with new
types of Josephson junctions  capable of reaching the
cosmologically
interesting frequency $\nu_c$ \cite{warburton}. 

The central hypothesis that we explore in this paper is that
vacuum fluctuations are {\em gravitationally active} (and hence
contribute to the dark energy density of the universe) if and
only if they are {\em measurable} (in the form of a spectral
density) in a suitable macroscopic or mesoscopic detector. We
will show that this basic hypothesis: a) provides a possible
solution to the cosmological constant problem; b) predicts the
correct order of magnitude of dark energy currently observed in
the universe; and c) is testable in future laboratory experiments
based on Josephson junctions. We also argue that the optimal
detector for measurable quantum noise spectra will typically
exploit macroscopic quantum effects in superconductors. From  our
measurability assumption we obtain a formula for the observable
dark energy density in the universe that is a kind of analogue of
the Stefan-Boltzmann law for vacuum energy, but in which the
temperature $T$ is not a free parameter but rather given by the
largest possible critical temperature $T_c$ of high-$T_c$
superconductors.

Our central hypothesis implies that vacuum fluctuations at very
high frequencies do not contribute to the cosmological constant.
In that sense it puts a cosmological limitation on general
relativity. However, it should be clear that all models of dark
energy in the universe require new physics in one way or another
\cite{copeland}. The advantage of our approach is that it is
experimentally testable.

This paper is organized as follows. In Section II we explain the
central hypothesis of this paper and discuss how it can help to
avoid the cosmological constant problem. In Section III we show
that assuming the central hypothesis is true one obtains the
correct order of magnitude of dark energy density in the
universe. Section IV shows how the near-equilibrium condition of
the fluctuation dissipation theorem can be used to obtain further
constraints on the dark energy density. Finally, in Section V we
provide some theoretical background for our measurability
approach. We discuss the fluctuation dissipation theorem and its
potential relation to dark energy, as well as the AC Josephson
effect and its relation to measurability of vacuum fluctuation
spectra.

\section{Measurability and gravitational activity of vacuum fluctuations}

Let us once again state the following {\bf central hypothesis}:

\begin{quote}

Vacuum fluctuations are {\em gravitationally active} if and only
if they are {\em measurable} in terms of a physically relevant
power spectrum in a macroscopic or mesoscopic detector

\end{quote}

It is important to make our usage of terms precise.
    \ben
    \item By  `vacuum
fluctuations that are gravitationally active' we mean those that
contribute to the currently measured dark energy density of the
universe.
    \item By `measurable vacuum fluctuations' we mean those
that induce a measurable quantum noise power spectrum in a
suitable macroscopic or mesoscopic  detector, for example in a
resistively shunted Josephson junction.
    \een

The central hypothesis is testable experimentally, e.g. in the
experiment by Warburton, Barber and Blamire \cite{warburton}
currently under way. If the central hypothesis is true, then the
vacuum fluctuations producing the measured spectra in the Josephson
junction are gravitationally active (i.e.\ contributing to dark
energy density), and hence there will be a cutoff near $\nu_c$ in
the measured spectrum.  Otherwise the dark energy density of the
universe is exceeded. The converse is also true: If the cutoff is
not observed, the observed vacuum fluctuations in the Josephson
junction cannot be gravitationally active.  In  this case the
central hypothesis is false.

We now demonstrate how, under the assumption that the central
hypothesis is true, the cosmological constant problem is solved
and at the same time the correct order of magnitude of dark energy
density in the universe is predicted.

Recall  that quantum field theories predict a divergent (infinite)
amount of vacuum energy given by
\begin{equation}
\hat{\rho}_{vac}= \frac{1}{2}(-1)^{2j} (2j+1)
\int_{-\infty}^\infty \frac{d^3k}{(2\pi)^3} \sqrt{\vec{k}^2+m^2}
\label{hatty}
\end{equation}
in units where $\hbar = c =1$. Here $\vec{k}$ is the momentum
associated with a vacuum fluctuation, $m$ is the mass of the
particle under consideration, and $j$ is the spin.  The central
hypothesis now immediately explains why most of the vacuum energy
in eq.~(\ref{hatty}) is not gravitationally active: In most cases
there will be no suitable detector to measure the vacuum
fluctuations under consideration. And what is not measurable is
also not observable. Only in very rare cases will there  be such
a detector, and only in these cases the corresponding vacuum
energy is measurable and thus physically relevant. According to
our hypothesis, the detectable part of vacuum energy is
gravitationally active and responsible for the current
accelerated expansion of the universe.

For strong and electro-weak interactions it is unlikely that a
suitable macroscopic detector exists that can measure the
corresponding vacuum spectra. The only candidate where we know that
a suitable macroscopic detector exists is the electromagnetic
interaction. Photonic vacuum fluctuations induce measurable spectra
in superconducting devices, as experimentally confirmed by Koch et
al. \cite{koch} up to frequencies of 0.6 THz. For photons, $m=0$ and
the integration over all $\vec{k}$ in eq.~(\ref{hatty}) is just an
integration over all frequencies $\nu$ since
$E=\sqrt{\vec{k}^2+m^2}=|\vec{k}|=h\nu =\hbar \omega$.

These detectors of photonic vacuum fluctuations will no
longer function if the frequency $\nu$ becomes too large. To see
this, consider the spectrum
\begin{equation}
S(\omega)= \frac{1}{2} \hbar \omega +\frac{\hbar
\omega}{\exp(\hbar \omega/kT)-1} \label{tommilein}
\end{equation}
occurring in the square brackets of eq.~(\ref{eq:power}). This
spectrum is at the root of the problem considered here and it is of
relevance for other mesoscopic systems as well \cite{mohanty}. The
spectrum in eq.~(\ref{tommilein}) is identical (up to a factor
$4/R$) to the measured spectrum in Josephson junctions. It arises
out of the fluctuation dissipation theorem \cite{welton,kubo,landau}
in a universal way and it formally describes a quantum mechanical
oscillator of frequency $\omega$. The linear term in $\omega$
describes the zeropoint energy of this quantum mechanical
oscillator, whereas the second term describes thermal states of this
oscillator. There is vacuum energy associated with the zeropoint
term and we may identify it with the source of dark energy (for more
details on the theoretical background, see Section 5). For the
vacuum fluctuation noise term $\frac{1}{2} \hbar \omega$ to be
measurable, it must be not too small relative  to the thermal noise
term $\hbar \omega/(e^{\hbar \omega/kT}-1)$. For low frequencies the
thermal noise dominates, for large frequencies the quantum noise.
The frequency $\omega_0$ where both terms have the same size is
given by
    \be
    \omega_0 = \frac{kT}{\hbar} \ln 3.
    \label{1}
    \ee
Now suppose we want to measure vacuum fluctuation spectra at very
large frequencies $\omega$. From eq.~(\ref{1}) we can achieve
higher frequencies by choosing to do our measurements at higher
temperatures since increasing the temperature increases the
frequency $\omega_0$. As  $\omega_0$ increases, the vacuum
fluctuation term dominates relative  to thermal noise for all
frequencies $\omega > \omega_0$.

There is however a practical limit to this procedure.  Namely, if
the temperature $T$ becomes too high, then a superconducting state
will no longer exist. Since superconducting devices such as
Josephson junctions appear to be the only experimentally feasible
devices to measure high frequency quantum noise spectra (see the
arguments in section 5), this means that there is a maximum
frequency $\omega_c$ above which a superconducting detector is no
longer functional  and the quantum fluctuation spectrum becomes
unmeasurable. This critical frequency is given by
\begin{equation}
\hbar \omega_c \sim kT_c, \label{2}
\end{equation}
where $T_c$ is the largest possible critical temperature of any
superconductor. Currently, the largest $T_c$ known for high-$T_c$
superconductors is approximately $T_c=139$ K \cite{high-Tc}.

In fact, technically feasible solutions of superconducting materials
that are used in practice to build Josephson junctions have lower
$T_c$. For example, the well-known YBCO materials have a maximum
critical temperature $T_c$ of 93 K \cite{tinkham}. To optimize a
Josephson  quantum noise detector one needs to avoid quasiparticle
currents, which means that the junction should operate at a
temperature $T_c'$ well below $T_c$. Let us choose as a rough
estimate $T_c'\approx 80$ K. One obtains $\nu_c'\sim kT_c'/h \approx
1.7$ THz. It is encouraging that this value is so close to that of
eq.~(\ref{eq:cutoff}), thus providing us with the correct order of
magnitude of measurable dark energy density if the central
hypothesis is true.

\section{Estimate of observable dark energy density}

To get a better estimate of the proportionality constant in
eq.~(\ref{2}), let us more carefully analyze when an experiment
based on Josephson junctions will be able to resolve quantum
noise. In  the data in Fig.~1,  the overall precision by which
the power spectrum can be measured at a given frequency in the
Koch experiment is of the order 10-40\%. Other experiments based
on SQUIDs yield fluctuations of similar order of magnitude in the
measured power spectrum \cite{asta}. The quantum noise term
$\frac{1}{2}\hbar \omega$ is measurable as soon as it is larger
than the standard deviation of the fluctuations of the measured
spectrum. That is to say,
\begin{equation}
\frac{1}{2} \hbar \omega > \eta \cdot \frac{\hbar
\omega}{e^{\hbar \omega / kT}-1} \label{3}
\end{equation}
where we estimate $\eta \simeq 0.1\,\to \, 0.4$ from the
fluctuations of the Koch data in Fig.~1. Condition (\ref{3}) can
be written as
\begin{equation}
\frac{\hbar \omega}{kT} >\ln (1+2\eta).
\end{equation}
We thus obtain from eq.~(\ref{2}) the critical frequency $\nu_c$
beyond which measurements of the spectrum become impossible as
\begin{equation}
\nu_c \approx 
\frac{\ln (1+2\eta)}{2\pi} \frac{kT_c}{\hbar}.
\end{equation}
The largest critical temperature of a high-$T$ superconductors
achieved so far is approximately $T_c=139$K \cite{high-Tc}.
Our result yields, with $T_c=139$K and $\eta \simeq 0.1\,\to \,
0.4$,
\begin{equation}
\nu_c\approx \ln (1+2\eta) \times 2.89 \times  10^{12} \mbox{Hz}
\simeq 0.5\, \to \, 1.7 \mbox{THz}.
\end{equation}

Recall that the dark energy density  measured in astronomical
observations is correctly reproduced for $\nu_c=1.69 \times
10^{12}$ Hz \cite{beck-mackey}. It is interesting that  our
argument based on measurability of vacuum fluctuations predicts
the correct order of magnitude of dark energy density in the
universe. This is especially hearting since the cosmological
constant problem  is usually plagued by estimates of vacuum energy
too large by a factor $10^{120}\,\,$\cite{weinberg,peebles}. Our
proposition for the resolution of  the cosmological constant
problem is simply that for frequencies higher than $\nu_c$ a
macroscopic detector no longer exists to measure the spectrum of
vacuum fluctuations. As soon as the vacuum fluctuations are no
longer measurable, by our central hypothesis they also can no
longer have a large-scale gravitational effect.

From this perspective, the dark energy density of the universe is
given by integrating the energy density over all {\em measurable}
vacuum fluctuations, to obtain
\begin{eqnarray}
\rho_{dark}&=&\int_0^{\nu_c} \frac{8\pi \nu^2}{c^3}
\frac{1}{2}h\nu d\nu \\
&=&\frac{\pi h}{c^3} \nu_c^4 \\
&=& \frac{\ln^4 (1+2\eta)}{8\pi^2} \frac{(kT_c)^4}{\hbar^3 c^3}.
 \label{b-m}
\end{eqnarray}
This result can be considered an analogue of the Stefan-Boltzmann
law
\begin{equation}
\rho_{rad}= \frac{\pi^2}{15} \frac{(kT)^4}{\hbar^3c^3}
\end{equation}
for radiation energy density.  The difference is that in
eq.~(\ref{b-m}) the temperature is  the largest possible critical
temperature $T_c$ of superconductors, and  the proportionality
constant $\frac{\ln^4 (1+2\eta)}{8\pi^2} \approx 1.4 \times
10^{-5}\, \to \, 1.5 \times 10^{-3}$ is much smaller than for the
Stefan-Boltzmann law (i.e. compared with $\frac{\pi^2}{15}\simeq
0.66$). Using $T_c=139$ K and the known dark energy density
$\rho_{dark}=(3.9\pm 0.4)\,\, \mbox{GeV/m$^3$}$ \cite{bennett,
spergel, spergelnew} we may write eq.~(\ref{b-m}) as
\begin{equation} \rho_{dark}=\sigma \frac{(kT_c)^4}{\hbar^3 c^3}
\end{equation}
where the dimensionless constant $\sigma$ is given by
\begin{equation}
\sigma = \frac{\ln^4 (1+2\eta)}{8\pi^2} =\hbar^3 c^3
\frac{\rho_{dark}}{(kT_c)^4}=(1.46\pm 0.15) \times 10^{-3}.
\end{equation}

\section{Further constraints on dark energy density}

The explanation of  why vacuum fluctuations produce a measurable
spectrum of noise in resistively shunted Josephson junctions is
based on two fundamental effects, the fluctuation dissipation
theorem \cite{kubo,kogan, welton, senitzky, landau} and the AC
Josephson effect \cite{josephson, tinkham, landau2}. The Josephson
effect requires the existence of a superconducting state, and in
this way we were led to an estimate on observable (measurable)
dark energy density in the previous section. The fluctuation
dissipation theorem requires that the system under consideration
is close to thermal equilibrium. We now show that the latter
condition also provides an upper bound on observable (measurable)
dark energy density. In other words, both effects imply that the
cosmological constant is small.

 The fluctuation dissipation theorem predicts a current
$(I)$ spectrum (A$^2$/Hz) in the shunt resistor given by
\begin{equation}
S_I(\omega) =2 \hbar \omega \coth \left( \frac{\hbar \omega}{2kT}
\right) Re \; Z^{-1} (\omega),
\end{equation}
and a voltage $(V)$  spectrum (V$^2$/Hz) given by
\begin{equation}
S_V(\omega) =2 \hbar \omega \coth \left( \frac{\hbar \omega}{2kT}
\right) Re \; Z (\omega),
\end{equation}
where $Z(\omega)$ is the impedance. As said before, the
derivation of the fluctuation dissipation theorem is based on the
assumption of thermal equilibrium in the resistor, or that the
system is at least near thermal equilibrium.

The noise spectra measured with Josephson junctions correspond to
real currents of real electrons, rather than virtual fluctuations,
and hence these currents will generate heat in the shunting
resistor. The dissipated power is $P=I \times  V$, which again
provides a reason why there must be an upper cutoff in the
measurable spectrum. Namely, if the frequency becomes too  high,
then the associated heat production through the power dissipation
will become  so large that it takes the system away from thermal
equilibrium.  Consequently,  the fluctuation dissipation theorem
would no longer  be valid.

To illustrate this, assume there is  a physical device that could
measure the vacuum noise spectra to frequencies much higher than
$\nu_c$, say of the same order of magnitude as relevant for the
Casimir effect.



Consider two Casimir plates separated by a distance $L$.  The
wavelengths of vacuum fluctuations that will fit into the cavity
formed by the plates must satisfy $\lambda \leq L$ but since
$\lambda = c/\nu$ this means that the minimum frequency of vacuum
fluctuations associated with the Casimir effect must satisfy
$$
\nu_{min} \geq \frac{c}{L}.
$$
Experimental measurements of the Casimir effect are  made with $L
\simeq O(0.1 \,\mu) \to O(1 \,\mu)$ \cite{bressi}, so   assume $L
= 1 \, \mu = 10^{-6}$ m to give $\nu_{min} \simeq 3 \times
10^{14}$ Hz, some $200$ times greater than the cutoff frequency
$\nu_c$ in eq.~(\ref{eq:cutoff}). If the predicted cutoff
frequency in the Josephson junction were increased to
$\nu_{min}$, this would imply an energy density of about
    \bea
    \rho_{dark} \left ( \frac{\nu_{min}}{\nu_c} \right ) ^4 &\simeq& 3.9
    \times 2^4 \times 10^8 \quad \mbox{GeV/m$^3$}\\ &=& 6.2 \times 10^9
    \quad \mbox{GeV/m$^3$}
    \eea
which is dissipated in the shunt resistor. Suppose that this energy
density came from a black body described by the Stefan-Boltzmann
law, i.e.
$$
\frac{\pi^2k^4}{15\hbar^3c^3} T^4 = 6.2 \times 10^{9} \quad
\mbox{GeV/m$^3$}.
$$
What would the corresponding temperature be? Quite high, namely $T
\simeq 6000$ K. It is clear that the Josephson experiment
could not operate with a heat source at that temperature, and
thermal equilibrium would be destroyed long before. This simple
argument once again shows that there must be a cutoff in the
measurable spectrum at frequencies much smaller than $\nu_{min}$.

If we make the same  estimate for the dark energy density, the
corresponding temperature of a black body with the same energy
density as dark energy density is $T \simeq
30$ K, so with some suitable cooling this neither
disturbs thermal equilibrium, nor disturbs a junction operating
in a superconducting state.

Note that the experiments that successfully confirm the QED
predictions of the Casimir effect (see, e.g., \cite{bressi}) just
provide measurements of the Casimir force, they do not provide us
with a measured spectrum of vacuum fluctuations, as required by the
central hypothesis. If the central hypothesis is correct, it is
clear that the vacuum fluctuations that formally (i.e. by entering
into the QED calculations) influence the Casimir effect are
gravitationally inactive since their frequency is much larger than
$\nu_c$ (see also \cite{padman2}).

\section{Theoretical background}

\subsection{The fluctuation dissipation theorem and dark energy}

Consider an arbitrary observable $x(t)$ of a quantum mechanical
system and consider the quantum mechanical expectation of the time
derivative of $x$, denoted by $\langle \dot{x} (t) \rangle$. Let
$F(t)$ be an external force. If linear response theory is applicable
we may write
\begin{equation}
\langle \dot{x} (t) \rangle = \int_{-\infty}^t G(t-t')F(t')dt',
\end{equation}
where the function $G$ is the `generalized conductance'. Its
Fourier transform is given by
\begin{equation}
G(\omega) = \int_0^\infty dt e^{i\omega t} G(t).
\end{equation}
The fluctuation dissipation theorem \cite{welton, senitzky, kubo,
kogan, landau} yields a very general relation between the power
spectrum $S_{\dot{x}}$ of the stochastic process $ \dot{x} (t)$
and the real part of $G(\omega)$:
\begin{eqnarray}
S_{\dot{x}} (\omega) &=& 2 \hbar \omega \coth \left( \frac{\hbar
\omega}{2kT}\right) Re \; G(\omega ) \\
&=&\left[ \frac{1}{2} \hbar \omega + \frac{\hbar\omega}{\exp
(\hbar\omega/kT)-1} \right] \cdot 4 Re \; G(\omega) \label{FDT}
\end{eqnarray}
The function  in square brackets is {\em universal}, it does not
depend on details of the quantum system considered, in particular on
its Hamiltonian $H$.  However,  $G(\omega )$ is system dependent.
The universal function in the square brackets can be physically
associated with the mean energy of a quantum mechanical oscillator
of frequency $\omega$ at temperature $T$. Its ground state energy is
given by $\frac{1}{2}\hbar \omega$. Note that this oscillator
has
nothing to do with the original Hamiltonian $H$ of the quantum
system under consideration. Also note that the fluctuation
dissipation theorem is valid for arbitrary Hamiltonians $H$, $H$
need not describe a harmonic oscillator at all but can be a much
more complicated Hamiltonian function. Nevertheless, the universal
function $H_{uni}=\frac{1}{2}\hbar \omega +\hbar\omega /(\exp (\hbar\omega / kT)-1)$
that occurs in the square brackets of eq.~(\ref{FDT}) can
always be formally interpreted as  the mean energy of a harmonic
oscillator.

Our physical interpretation is to identify the zeropoint energy of
$H_{uni}$ as the source of dark energy. This
oscillator is universally present everywhere, but typically
manifests  measurable effects only in dissipative media.
Note that the term $\frac{1}{2}\hbar \omega$, which describes
the zeropoint energy of this oscillator, is invariant under
transformations $H\to H+const$ of the system Hamiltonian $H$. This
is obvious, since the fluctuation dissipation theorem is valid for
{\em any} Hamiltonian $H$. In other words, the zeropoint energy of our
dark energy oscillator $H_{uni}$ is an invariant, it is invariant under
renormalization of the system Hamiltonian $H$. This is a strong hint
that the corresponding vacuum energy $\frac{1}{2}\hbar \omega$ is
indeed a physically observable quantity. It is likely to have
gravitational relevance since it is invariant under arbitrary
re-definitions of the system Hamiltonian $H$.

A  further interpretation (as suggested in the classical papers and
textbooks on the subject \cite{welton, senitzky,gardiner}) is to
associate the term $\frac{1}{2}\hbar \omega$ with the zeropoint
fluctuations of the electromagnetic field.
We thus arrive at the following, in our opinion, most plausible
interpretation of the fluctuation dissipation theorem: The term
$\frac{1}{2}\hbar \omega$ describes the {\em gravitationally active}
part of the vacuum fluctuations of the electromagnetic field.

Once again let us emphasize that we consider
the universal function $H_{uni}$ occuring
in the fluctuation dissipation theorem as
a potential source of dark energy, rather than the system
dependent Hamiltonian $H$.
In recent papers \cite{jetzer,doran}, the role of
$H$ and $H_{uni}$ is confused.
The authors re-derive
the well-known fact that the fluctuation dissipation theorem is
valid for arbitrary Hamiltonians $H$, in particular for those
where an arbitrary additive constant is added to $H$. However,
their argument
relates to the system Hamiltonian $H$ and not to $H_{uni}$
and is thus not applicable to our model of dark energy
based on $H_{uni}$.
In particular, the considerations in \cite{jetzer,doran}
provide no insight into the physical
interpretation of the vacuum energy
associated with $H_{uni}$, which is invariant and universal.

\subsection{The AC Josephson effect and measurability
of vacuum fluctuations}

The fluctuation dissipation theorem quite generally explains the
existence of quantum noise in resistors, but on its own it does not
suffice to make the power spectrum of quantum fluctuations
measurable in an experiment. Putting a voltmeter directly into the
dissipative medium would not prove to be a feasible method to
measure the zeropoint spectrum at high frequencies. Rather, we need
a more sophisticated method, and the AC Josephson effect
\cite{josephson, tinkham, landau2} satisfies this requirement.

To briefly explain this effect, remember that a Josephson junction
consists of two superconductors with an insulator sandwiched in
between. In the Ginzburg-Landau theory, each superconductor is
described by a complex wave function, whose absolute value squared
yields the density of superconducting electrons. Denote  the phase
difference between the two wave functions of the two superconductors
by $\varphi (t)$. Josephson \cite{josephson} made the remarkable
prediction that at zero external voltage a superconductive current
given by
\begin{equation}
I_s=I_c \sin \varphi \label{25}
\end{equation}
flows between the two superconducting electrodes. Here $I_c$ is
the maximum superconducting current the junction can support.
Moreover, he predicted that if a voltage difference $V$ is
maintained across the junction, then the phase difference
$\varphi$ evolves according to
\begin{equation}
\dot{\varphi}=\frac{2eV}{\hbar},
\end{equation}
i.e.\ the current in eq.~(\ref{25}) thus becomes an oscillating current with
amplitude $I_c$ and frequency
\begin{equation}
\nu = \frac{2eV}{h}. \label{freq}
\end{equation}
This frequency is the well-known Josephson frequency, and the
corresponding effect is called the AC Josephson effect. The quantum
energy $h\nu$ given by eq.~(\ref{freq}) can be interpreted as the
energy change of a Cooper pair that is transferred across the
junction. The AC Josephson effect is a very general and universal
effect that always occurs whenever two superconducting electrodes
are connected by a weak link.

The AC Josephson  effect connects quantum mechanics (i.e.
differences of phases of macroscopic wave functions) with measurable
classical quantities (currents or voltages). A Josephson junction
can be regarded as a perfect voltage-to-frequency converter,
satisfying the  relation $2eV=h\nu$. For distinct DC voltages, it is
also a perfect frequency-to-voltage converter. This inverse AC
Josephson effect is, for example, used to maintain the SI unit Volt.

In the experiments of Koch et al. \cite{koch} the quantum noise in
the shunt resistor is mixed down at the Josephson frequency
$2eV/h$ to produce measurable voltage fluctuations. The
measurement frequency in these experiments is usually much
smaller than the Josephson frequency. However, due to the
specific nonlinear properties of Josephson junctions, the
measured voltage fluctuations are influenced by quantum
fluctuations at the Josephson frequency and its harmonics
\cite{koch80theory, levinson}. In this way quantum fluctuations
in the THz regime become experimentally accessible. The frequency
variable $\nu$ in Fig.~1 is experimentally varied by applying
different DC voltages across the junction, thus making direct use
of formula (\ref{freq}). Josephson oscillations are clearly
necessary to do these types of measurements.

The AC Josephson effect has been experimentally observed up to
Josephson frequencies in the low-THz region.  The energy gap in
cuprates limits the maximum value of the Josephson frequency to $\nu
\sim 15$ THz, but in practice one seems to be able to only reach the
2 THz region \cite{stepan}. If the central hypothesis is true, then
the largest attainable  Josephson frequency also constrains the dark
energy density of the universe.

\section{Conclusions and outlook}

In this paper we have proposed a possible solution to the
cosmological constant problem, based on the central hypothesis that
vacuum fluctuations are only {\em gravitationally active} if they
are {\em measurable} in the form of a noise spectrum in suitable
macroscopic or mesoscopic detectors. From this assumption a
universal low-energy cutoff frequency $\nu_c$ consistent with the
currently observed dark energy density in the universe is predicted
in a rather natural way. One obtains $\nu_c \sim kT_c/h$, where
$T_c$ is the largest possible critical temperature of high-$T_c$
superconductors. This means $\nu_c$ is in the THz region. Moreover,
the dark energy density of the universe is most naturally identified
as ordinary electromagnetic vacuum energy of virtual photons with
frequency $\nu <\nu_c$, and given by a kind of analogue of the
Stefan-Boltzmann law for dark energy, $\rho_{dark} \sim
(kT_c)^4/(\hbar^3 c^3)$.

Suppose  the universal cutoff $\nu_c= 1.7$ THz corresponding to
the dark energy density is observed in the experiment planned by
Warburton et al. \cite{warburton}. Would this invalidate
successful QED predictions for other effects such as the Casimir
effect \cite{bressi, padman2} or Lamb shift?

We think the answer is `no'. Once again we  emphasize that the
cutoff is predicted for the {\em measurable} spectrum. Virtual
photons as a useful tool for the theoretician are still allowed to
persist at higher frequencies. These can then still be used as a
theoretical tool to do calculations for the Casimir effect, Lamb
shift, spontaneous emissions of atoms etc. in just the same way as
before, keeping in mind that in many cases, e.g.\ for the Casimir
effect, they are not needed  at all to explain the effect
\cite{jaffe}.

The central hypothesis merely implies that for photonic vacuum
fluctuations the {\em measurability} in the form of a physically
relevant spectrum ceases to exist for  $\nu > \nu_c = 1.7$ THz, for
the reasons we have given above. This is connected with a kind of
{\em phase transition} for the gravitational activity of the virtual
photons at $\nu=\nu_c$. It is indeed plausible that vacuum
fluctuations are only gravitationally active if they are measurable
in the form of a frequency spectrum in a macroscopic or mesoscopic
detector, as stated by our central hypothesis. How else should or
could these vacuum fluctuations push galaxies apart and accelerate
the expansion of the universe if the effect is not measurable with a
macroscopic detector? As shown above the central hypothesis leads to
the correct dark energy density in the universe, and the
cosmological constant problem is avoided. The validity or
non-validity of the central hypothesis will soon be tested
\cite{warburton}.



\begin{acknowledgments}
This work was supported by the Engineering and Physical Sciences
Research Council (EPSRC, UK), the Natural Sciences and Engineering
Research Council (NSERC, Canada) and the Mathematics of
Information Technology and Complex Systems (MITACS, Canada). This
research was carried out while MCM was visiting the Mathematics
Institute of the University of Oxford.
\end{acknowledgments}


\end{document}